# Serotonergic and noradrenergic contributions to human motor cortical and spinal motoneuronal excitability

Jacob R. Thorstensen[1], Tyler T. Henderson[2] and Justin J. Kavanagh[2]

[1]School of Biomedical Sciences, The University of Queensland, Brisbane, Australia
[2]Menzies Health Institute Queensland, Griffith University, Gold Coast, Australia

**Abstract**

Animal models indicate that motor behaviour is shaped by monoamine neurotransmitters released diffusely throughout the brain and spinal cord. We present strong evidence that human motor pathways are equally affected by neuromodulation through noradrenergic and serotonergic projections arising from the brainstem. To do so, we have identified and collated human experiments examining the off-label effects of well-characterised serotonergic and noradrenergic drugs on lab-based electrophysiology measures of corticospinal-motoneuronal excitability. Specifically, we focus on the effects that serotonin and noradrenaline associated drugs have on muscle responses to magnetic or electrical stimulation of the motor cortex and peripheral nerves, and other closely related tests of motoneuron excitability, to best segment drug effects to a supraspinal or spinal locus. We find that serotonin enhancing drugs tend to reduce the excitability of the human motor cortex, but that augmented noradrenergic transmission increases motor cortical excitability by enhancing measures of intracortical facilitation and reducing inhibition. Both monoamines tend to enhance the excitability of human motoneurons. Overall, this work details the importance of neuromodulators for the output of human motor pathways and suggests that commonly prescribed monoaminergic drugs have off-label motor control uses outside of their typical psychiatric/neurological indications.

# 1. Introduction

Excitatory synaptic input from cortical motor pathways directly activates spinal motoneurons, often via interneurons, which cause motoneurons to discharge and activate muscle fibres. Two main neurotransmitter mechanisms control the firing properties of cells within the motor cortex and motoneurons (Heckman and Enoka, 2012; Heckman *et al.*, 2009; Rekling *et al.*, 2000; Vitrac and Benoit-Marand, 2017). In general, fast acting ionotropic inputs, such as corticospinal neurons, afferents projecting from muscles and joints, or intracortical and spinal interneurons, directly open and close ion channels to depolarize/hyperpolarize their post-synaptic targets. In contrast, metabotropic inputs do not directly cause excitation or inhibition of individual neurons but rather modify their responsiveness to ionotropic inputs. Put simply, ionotropic neurotransmitters activate neurons, whereas metabotropic neurotransmitters set the gain or excitability. Metabotropic projections (i.e., cells that release monoamine neurotransmitters) arise from the brainstem and project diffusely throughout the brain and spinal cord, and these projections are commonly labelled as neuromodulatory (note: this is a physiological process and is not to be confused with 'neuromodulatory' technologies). There is no principal neurotransmitter system solely responsible for the neural control of movement, and the net excitability of motor circuits in the brain and spinal cord are controlled by a complex balance of excitatory, inhibitory and neuromodulatory inputs.

Serotonin (5-HT) and noradrenaline (NA) exert several key neuromodulatory effects within motor circuits of the brain and spinal cord of animals (Heckman *et al.*, 2008; Heckman *et al.*, 2009; Perrier and Cotel, 2015; Perrier *et al.*, 2013; Vitrac and Benoit-Marand, 2017). The human motor cortex is innervated by ascending serotonergic and noradrenergic projections from the raphe nuclei and locus coeruleus respectively (Gaspar *et al.*, 1989; Javoy-Agid *et al.*, 1989; Raghanti *et al.*, 2008), and *in vivo* animal studies indicate that 5-HT and NA increase the excitability of the motor cortex (Schiemann *et al.*, 2015; Scullion *et al.*, 2013). Motoneurons are the final output neurons for all motor behaviour and are also innervated by 5-HT and NA neurons descending from the brainstem (Alvarez *et al.*, 1998; Rajaofetra *et al.*, 1992). 5-HT and NA tend to increase the excitability of motoneurons and have been hypothesised to operate as gain controllers for muscle activation at a spinal level (Johnson and Heckman, 2014), thus enabling the scaling of motoneuron excitability to motor behaviour. In this light, computer simulations generated with data obtained from animal motoneuron experiments indicate that muscles would not even produce half of their maximal force in the absence of metabotropic inputs, even when ionotropic input is maximal (Heckman *et al.*, 2008).

Most of the work concerning neuromodulators, and their effects on motor circuits in the brain and spinal cord, have involved slice preparation experiments. However, using mechanisms derived from *in vitro* work to explain neurophysiological processes within the intact human nervous system is not a straightforward process, whereby the inputs to and voltage properties of individual neurons cannot be directly measured or controlled. To better determine whether results from reductionist preparation experiments align with human motor behaviour, several 5-HT and NA specific drugs have been used off-label as a tool to target the activity of neuromodulators or their associated receptors, and post-drug changes in motor behaviour subsequently studied and compared to a placebo or drug free condition. Alternatively, many clinical studies use monoaminergic drugs as an intervention for movement disorders and these studies are an untapped resource for understanding how neuromodulators affect human motor activity. In both situations, monoaminergic drugs either mimic or 'knockout' the typical physiological effects exhibited by 5-HT and NA in the nervous system, thus enabling causative links to be derived between monoamines and movement in intact human volunteers.

As orally consumed or intravenously applied drugs are non-selective to different structures in the brain and spinal cord, non-invasive neurostimulation techniques have also been used to activate the motor cortex and/or motoneurons to better segment the supraspinal and spinal effects of a given drug. Specifically, transcranial magnetic stimulation (TMS) of the motor cortex or the electrical/magnetic stimulation of corticospinal axons, and the electrical, mechanical, or vibratory activation of a peripheral nerve evokes responses in muscle that can be recorded with electromyography (EMG) (McNeil *et al.*, 2013; Monjo and Shemmell, 2020; Terao and Ugawa, 2002). The size of these evoked muscle responses provides a wealth of information regarding the excitability of the motor cortex, corticospinal pathway and motoneurons. When neurostimulation and other closely related techniques are paired with well-characterised drug interventions, better links can be made between human, animal, and *in vitro* mechanisms concerning neuromodulation.

There has not yet been an effort to amalgamate human studies involving both: i) the administration of 5-HT or NA drugs and ii) lab-based measures of corticospinal-motoneuronal excitability. In this review we gather and report human studies that have examined the effects of serotonergic and noradrenergic medications on muscle responses to magnetic or electrical stimulation of the motor cortex and peripheral nerves, and other closely related tests of motoneuron excitability (reflex responses to vibration or stretch, and EMG-derived estimates of motoneuronal persistent inward current amplitude, i.e., PIC amplitude). Our search strategy involved the entry of several keywords into PubMed and Google Scholar as well as cross-referencing identified papers. Specific keywords included drug classes (e.g., selective serotonin reuptake inhibitor/SSRI, amphetamine), common drug names (e.g., paroxetine, reboxetine), or neurotransmitter names (e.g., serotonin, noradrenaline/norepinephrine) alongside one or more terms relating to measurement techniques (e.g., transcranial magnetic stimulation, nerve stimulation, spinal reflex) and/or their associated outcomes (e.g., motor evoked potential, silent period, F wave). Using this search strategy, we identified 33 5-HT and 46 NA drug experiments, and an additional 11 experiments examining drugs with combined 5-HT/NA effects and summarise the results of these studies in tables (see Tables 1 to 3). Using what we know about the physiology of the abovementioned lab-based electrophysiology measures, and a synthesis of the results from the studies we identified, we speculate on how neuromodulators shape the output of the human brain and spinal circuits that activate muscle.

## 2. The serotonin system

*2.1 Serotonergic effects on resting muscle responses to motor cortical stimulation*

TMS studies indicate that 5-HT drugs have complex effects on the excitability of the motor cortex and corticospinal pathway in awake human subjects (Acler *et al.*, 2009; Batsikadze *et al.*, 2013; Bonin Pinto *et al.*, 2019; Busan *et al.*, 2009; Eichhammer *et al.*, 2003; Gerdelat-Mas *et al.*, 2005; Henderson *et al.*, 2022; Ilic *et al.*, 2002; Jeng *et al.*, 2020; Khedr *et al.*, 2020; Kuo *et al.*, 2016; Lagas *et al.*, 2016; McDonnell *et al.*, 2018; Melo *et al.*, 2021; Minelli *et al.*, 2010; Nitsche *et al.*, 2009; Pleger *et al.*, 2004; Robol *et al.*, 2004; Thorstensen *et al.*, 2021; Thorstensen *et al.*, 2020; Werhahn *et al.*, 1998) (Table 1). Motor cortical TMS activates the axons of corticospinal cells directly or via the activation of intracortical neurons that directly synapse onto these cells. Consequently, TMS of the motor cortex generates several descending volleys that arrive at spinal motoneurons, generating a compound muscle action potential that can be recorded with surface EMG at the muscle as a motor evoked potential (MEP). The physiology of TMS has been reviewed in detail elsewhere (Hallett, 2000; Terao and Ugawa, 2002; Wasserman *et al.*, 2008), however it is important to note that, the MEP provides an index of excitability for the whole motor pathway from cortex to muscle and is thus influenced by excitability changes at both the motor cortex and motoneurons (McNeil *et al.*, 2013).

On three occasions (Gerdelat-Mas *et al.*, 2005; Ilic *et al.*, 2002; Khedr *et al.*, 2020), single dose SSRI administration increased the size of MEPs obtained at rest (i.e., when there is no descending drive to the muscle). Yet longer-term SSRI administration was also shown to decrease the resting MEP (Gerdelat-Mas *et al.*, 2005), indicating that the number of SSRI administrations influences the direction of MEP size changes. Compared to single dose administration of an SSRI, the repeated administration over several weeks causes a de-sensitisation of 5-HT responsive inhibitory 5-HT$_1$ auto receptors and more 5-HT release (Stahl, 1998), potentially explaining the divergence of resting MEP findings between single and repeated doses of SSRI.

The stimulation intensity needed to reach resting motor threshold (RMT) is not consistently affected by serotonergic drugs (Busan *et al.*, 2009; Eichhammer *et al.*, 2003; Gerdelat-Mas *et al.*, 2005; Ilic *et al.*, 2002; Lagas *et al.*, 2016; McDonnell *et al.*, 2018; Pleger *et al.*, 2004; Werhahn *et al.*, 1998), except for a small minority of studies whereby SSRI medications increase threshold (Acler *et al.*, 2009; Minelli *et al.*, 2010; Robol *et al.*, 2004). RMT is typically defined as the intensity needed to elicit a MEP of 50 µV peak-peak amplitude in at least 50% of stimulations (Groppa *et al.*, 2012; Rossini *et al.*, 2015). Stimulation intensities at RMT activate the cortico-cortical and corticospinal projections most easily excited by the magnetic pulse, and most likely lower-threshold spinal motoneurons if descending volley/s are evoked. Hence, the absence of clear and consistent 5-HT drug effects on RMT indicates that serotonergic projections to the motor cortex may not affect lower threshold inputs to motoneurons (regarding recruitment via TMS, which may not recruit the same inputs to motoneurons as voluntary activity), and/or the excitability of smaller motoneurons usually activated by lower levels of excitatory synaptic input. Findings may also indicate that lower-intensity TMS assesses excitability components of interneurons and/or corticospinal cells that are not strongly mediated by neuromodulators (e.g., axonal excitability or the magnitude of pre-synaptic glutamate release). Likewise, compared to suprathreshold stimulation (i.e., stimulation 120% RMT and greater) recruiting a large sample of cells into the resting MEP, the excitability of a smaller sample of the corticospinal-motoneuronal pathway was assessed for RMT determination, and TMS may not be sensitive enough to pick up excitability changes under these conditions.

*2.2 Intracortical facilitation and inhibition are modulated by 5-HT drugs*

When conditioned by a preceding stimulation to the same site of the motor cortex, the amplitude of the resting MEP provides an index of intracortical inhibitory or facilitatory strength (Hallett, 2000; Kujirai *et al.*, 1993; Rossini *et al.*, 2015; Terao and Ugawa, 2002; Wasserman *et al.*, 2008; Ziemann *et al.*, 1996). At very short interstimulus intervals (typically 2-3 ms), a subthreshold conditioning stimulus reduces the amplitude of a proceeding suprathreshold stimulation evoked test MEP, and this is called short interval intracortical inhibition (SICI). At an interstimulus interval of ~10 ms, the effect of a subthreshold conditioning stimulation increases the size of the MEP, and this is deemed intracortical facilitation (ICF). Long interval intracortical inhibition (LICI) is typically when two suprathreshold stimulations are separated by 100 ms. Regarding paired pulse TMS, several studies indicate that single dose SSRI administration decreases measures of facilitation and increases measures of inhibition (Eichhammer *et al.*, 2003; Gerdelat-Mas *et al.*, 2005; Ilic *et al.*, 2002; Minelli *et al.*, 2010; Robol *et al.*, 2004). Although it is tempting to conclude that 5-HT has a net effect of reducing intracortical excitability, separate intracortical circuits likely contribute to the conditioning of a MEP for ICI and ICF responses and these circuits do not necessarily overlap (Ziemann *et al.*, 1996). Importantly, the preceding short interval conditioning pulse that is used to reduce the size of the MEP preferentially activates low threshold inhibitory cells that generate inhibitory postsynaptic potentials (IPSPs) in excitatory cortical interneurons and/or corticospinal cells (Paulus *et al.*, 2008; Ziemann *et al.*, 2015). Hence, the enhanced availability of 5-HT with SSRI administration perhaps augments the magnitude and/or efficacy of inhibitory intracortical neurotransmission. Whatever the mechanism underlying SICI, the pharmacological activation of the $5-HT_{1B/1D}$ receptor reduces SICI (Werhahn *et al.*, 1998), indicating that this receptor may be involved. For ICF, enhanced 5-HT availability might decrease the excitability of an intracortical facilitatory circuit that is also activated by a subthreshold conditioning pulse to the motor cortex, so that the suprathreshold test pulse recruits less corticospinal neurons that are usually facilitated by these excitatory intracortical inputs.

In contrast to single dose SSRI administration, longer-term use of an SSRI was shown to increase ICF in healthy participants (Gerdelat-Mas *et al.*, 2005) and decrease LICI in individuals with depression (Jeng *et al.*, 2020), indicating that the direction of paired pulse MEP changes depend on the number of SSRI doses. In the unaffected hemisphere of individuals with stroke, chronic SSRI administration increases ICF (Bonin Pinto *et al.*, 2019), but both decreases and increases in SICI have been observed in separate studies (Acler *et al.*, 2009; Bonin Pinto *et al.*, 2019).

*2.3 5-HT drugs affect muscle responses to motor cortical stimulation during voluntary contraction*

Motor cortical TMS can be applied when the target muscle is voluntarily activated (i.e., to obtain an 'active' MEP, usually during brief isometric contractions). MEPs obtained during muscle contraction are larger than those obtained in resting muscle, as there is ongoing voluntary activation of the motor cortex and motoneurons (McNeil *et al.*, 2013). Accordingly, compared to rest, more motoneurons are in the subliminal fringe and closer to their firing thresholds, meaning that TMS is better able to recruit more motoneurons into the MEP and assess the excitability of a larger sample of efferent projections to muscle. When the muscle is pre-activated by voluntary activity, whereby voluntary contractions also probably cause some natural release of 5-HT to motoneurons (Jacobs *et al.*, 2002; Veasey *et al.*, 1995), SSRIs do not appear to affect the size of active MEPs in healthy participants (Henderson *et al.*, 2022; Thorstensen *et al.*, 2020). Yet, an isolated study investigating a single dose SSRI found an increase in active MEP size in individuals with depression (Khedr *et al.*, 2020). Perhaps, compared to non-depressed individuals, the active motor cortex and/or motoneuron pools in depressed individuals are more sensitive to 5-HT effects from the SSRI. Also of note, antagonism of the 5-HT$_2$ receptor decreases the size of MEPs obtained during muscle contractions in healthy participants (Thorstensen *et al.*, 2021). The active motor threshold (AMT), typically the stimulation intensity required to elicit a 200 μV peak-to-peak MEP during weak isometric contraction (in at least 50% of stimulations) (Groppa *et al.*, 2012; Rossini *et al.*, 2015), is not affected by 5-HT drugs (Busan *et al.*, 2009; Eichhammer *et al.*, 2003; Gerdelat-Mas *et al.*, 2005; Ilic *et al.*, 2002; Thorstensen *et al.*, 2021).

Motor cortical TMS can also be used to elucidate corticospinal-motoneuronal inhibitory mechanisms during muscle activity. Specifically, after the active MEP, a transient interruption to the voluntary activation of motoneurons can be observed with single pulse TMS to the motor cortex, and this is measured as a silent period in the ongoing EMG signal (Hupfeld *et al.*, 2020; Skarabot *et al.*, 2019). A longer duration silent period denotes more inhibitory activity within the motor cortex and/or at motoneurons, whereby the latter part of the silent period is likely attributed to long-lasting intracortical inhibition mediated by γ-aminobutyric acid (GABA) receptors (Paulus *et al.*, 2008; Ziemann *et al.*, 2015). Although paired pulse measures of inhibition likely reflect a different inhibitory mechanism/s than those that cause the TMS-induced silent period, we found that 5-HT mediated increases in paired pulse measures of inhibition are further corroborated by silent period findings across several studies, whereby single dose SSRI administration lengthens the TMS-induced silent period in healthy subjects (Henderson *et al.*, 2022; Robol *et al.*, 2004; Thorstensen *et al.*, 2020). In this regard, during voluntary contraction, enhanced 5-HT availability from the SSRIs could have caused a longer-lasting TMS-induced inhibition of corticospinal cells. A longer-lasting inhibition of motor cortical projections to motoneurons will cause a longer-lasting dis-facilitation of the motoneurons voluntarily recruited into contraction, resulting in the SSRI mediated lengthening of the TMS-induced silent period. Interestingly, antagonism of the 5-HT$_2$ receptor also increases silent period duration (Thorstensen *et al.*, 2021). In an isolated study (Busan *et al.*, 2009), repeated-dose (i.e., longer term) SSRI administration shortens the TMS-induced silent period, which provides further evidence that the direction of SSRI effects on TMS-induced muscle responses depend on the number of SSRI exposures.

It is important to emphasise that TMS of the motor cortex cannot discern between drug effects at different levels of the motor system, whereby 5-HT drugs are non-selective and likely also modulate the excitability of spinal cord motor circuits, and this can affect muscle responses to motor cortical stimulation. Consistent with TMS-induced silent period findings, the peripherally induced EMG silent period in a hand muscle (evoked through cutaneous electrical stimuli to the finger) is also lengthened by the single dose of an SSRI (Pujia *et al.*, 2014), potentially signifying 5-HT effects on inhibitory neurotransmission within the spinal cord.

*2.4 A complex role for 5-HT in the excitability of motoneurons to corticospinal stimulation and antidromic activation*

With TMS of the motor cortex, all evoked volleys that descend via corticospinal axons must pass through synapses with motoneurons and spinal interneurons before arriving at muscle fibres to be recorded as a MEP. Hence, the size of TMS-evoked muscle responses is heavily influenced by the excitability of sensorimotor circuits in the spinal cord. Considering this, tests of motoneuron excitability help elucidate whether changes to cortically evoked responses are due to supraspinal or spinal changes (McNeil *et al.*, 2013). Indeed, tests of motoneuron excitability indicate that 5-HT drugs have strong effects in the human spinal cord (D'Amico *et al.*, 2017; D'Amico *et al.*, 2013a; D'Amico *et al.*, 2013b; Gerdelat-Mas *et al.*, 2005; Goodlich *et al.*, 2023; Gourab *et al.*, 2015; Ilic *et al.*, 2002; Kamper *et al.*, 2022; Kavanagh *et al.*, 2019; Minelli *et al.*, 2010; Murray *et al.*, 2010; Pujia *et al.*, 2014; Robol *et al.*, 2004; Sattler *et al.*, 2000; Thorstensen *et al.*, 2022; Wei *et al.*, 2014) (Table 1).

Electrical or magnetic stimulation at the cervicomedullary junction, which evokes cervicomedullary motor evoked potentials (CMEPs) in upper-limb muscles, is the best spinal level control for MEPs as it activates many of the same inputs to motoneurons as TMS of the motor cortex (Taylor, 2006; Taylor and Gandevia, 2004). CMEPs have a strong monosynaptic component for proximal upper-limb muscles such as the biceps brachii (Petersen *et al.*, 2002). We only identified two studies which assessed the effects of 5-HT drugs on CMEPs. Importantly, for the resting biceps brachii, 5-HT$_2$ receptor antagonism does not affect CMEP size (Thorstensen *et al.*, 2022), but 5-HT$_{1A}$ receptor agonism does (D'Amico *et al.*, 2017). There are several key explanations that may explain these results. Regarding the somato-dendritic Gq coupled 5-HT$_2$ receptor, animal (Harvey *et al.*, 2006a, b; Murray *et al.*, 2011a; Perrier and Delgado-Lezama, 2005; Perrier and Hounsgaard, 2003) and human (D'Amico *et al.*, 2013b; Goodlich *et al.*, 2023) studies indicate that this receptor increases motoneuron excitability by facilitating PICs, thus greatly amplifying ionotropic input from descending or sensory sources, and causing the prolonged discharge of motoneurons. Yet in a resting motoneuron pool, the size of a CMEP is unlikely to reflect the contribution of dendritic PICs to motoneuron excitability (especially the component of the total PIC mediated by slowly activating Ca$^{2+}$ channels). This is because cervicomedullary stimulation probably only causes a brief excitatory postsynaptic potential (EPSP) in motoneurons but sustained excitatory input (lasting at least half a second in duration) is needed to cause repetitive firing of motoneurons and PIC activation (Li *et al.*, 2004). This contrasts with the Gi/Go coupled 5-HT$_{1A}$ receptor which is located at the axon initial segment (i.e., the site responsible for spike initiation) and shown to reduce motoneuron excitability in turtle preparations (Cotel *et al.*, 2013; Perrier *et al.*, 2017). In this regard, it seems most likely that agonism of 5-HT$_{1A}$ receptors reduces the size of CMEPs because less motoneurons can transmit the single spike past the initial segment and towards the muscle.

Considering the extra-synaptic location of the 5-HT$_{1A}$ receptor, it is not entirely clear if/when the 5-HT$_{1A}$ receptor is typically activated during human motor behaviour. 5-HT has a remarkable ability to increase the excitability of turtle, rat, and cat motoneurons (Perrier and Cotel, 2015; Perrier *et al.*, 2013), and the initial segment 5-HT$_{1A}$ receptor is perhaps only activated when the concentration of 5-HT in the synapse is sufficiently high, and 5-HT diffuses away from the synapse (Cotel *et al.*, 2013; Perrier *et al.*, 2017). There were no studies combining SSRI administration and cervicomedullary stimulation to investigate this possibility.

The supramaximal electrical stimulation of a peripheral motor nerve evokes antidromic spikes, which cause the recurrent orthodromic discharge of a small number of higher-threshold motoneurons, and these discharges are recorded with EMG in the resting muscle as F-waves (Fisher, 1992; McNeil *et al.*, 2013; Mesrati and Vecchierini, 2004). Motoneurons do not require synaptic excitation to produce an F-wave, making it a useful adjunct test of motoneuron excitability to complement cervicomedullary stimulation reflecting the synaptic activation of motoneurons. SSRIs do not affect F-wave amplitude or persistence (Gerdelat-Mas *et al.*, 2005; Ilic *et al.*, 2002; Kavanagh *et al.*, 2019; Minelli *et al.*, 2010; Robol *et al.*, 2004; Sattler *et al.*, 2000), indicating that enhanced synaptic 5-HT alone does not affect the recurrent discharge of motoneurons to antidromic activation. Yet, when F-waves are obtained directly after fatiguing muscle contractions in the presence of SSRI administration, which in combination likely cause strong 5-HT release into the spinal cord, they are smaller compared to placebo (Kavanagh *et al.*, 2019). This finding is consistent with the 5-HT 'spill over' hypothesis of central fatigue demonstrated in turtle preparations (Cotel *et al.*, 2013; Perrier *et al.*, 2017). In turtle preparations, prolonged stimulation of serotonergic projections to motoneurons results in a high concentration of synaptic 5-HT and causes 5-HT to diffuse away from the synapse and activate extra synaptic inhibitory 5-HT$_{1A}$ receptors (that reduce motoneuron spiking). In humans, the extra 5-HT from the SSRI likely causes more spill over and greater 5-HT$_{1A}$ receptor activation to decrease the likelihood of a recurrent discharge reaching the muscle to be recorded as an F-wave. In support of a 5-HT$_{1A}$ receptor locus for SSRI-mediated reductions in F-waves of fatigued muscle, the pharmacological activation of this receptor with the agonist buspirone also reduces F-waves (D'Amico *et al.*, 2017), but a reduction in F-waves has also been reported with 5-HT$_2$ antagonism (Thorstensen *et al.*, 2022).

*2.5 The excitability of motoneurons to sensory afferent input is modulated by 5-HT drugs*

Motoneurons can be reflexively recruited by the electrical or mechanical activation of sensory inputs in the periphery, and muscle responses to sensory afferent activation provide an index of excitability for sensorimotor circuits at the level of the spinal cord. For example, the electrical stimulation of Ia afferent axons provides a synchronous recruitment of motoneurons and elicits EMG recorded H-reflexes, the size of which may provide an index of motoneuron excitability to sensory input (Misiaszek, 2003; Theodosiadou *et al.*, 2023; Zehr, 2002). Activation of the 5-HT$_{1B/1D}$ receptor with an agonist reduces the size of H-reflexes in both healthy participants and in individuals with spinal cord injury (D'Amico *et al.*, 2013a). Yet, the electrical properties of Ia afferent fibres can modify sensory transmission to motoneurons (Rudomin and Schmidt, 1999; Stein, 1995), and a reduction in H-reflexes can be a consequence of pre-synaptic and not motoneuron excitability changes. In this light, sensory afferents express 5-HT$_1$ receptors that when activated, inhibit the output of segmental inputs to motoneurons (Murray *et al.*, 2011b). The 5-HT$_{1B/1D}$ receptor probably reduces the H-reflex through reductions in sensory transmission, whereby activation of this inhibitory receptor reduces the EPSP generated by Ia afferent and/or excitatory interneurons. Consequently, motoneurons are less likely to be recruited into the H-reflex, even though post-synaptic excitability is presumably unaffected by a 5-HT$_{1B/1D}$ agonist.

The effects of 5-HT drugs on motoneuron excitability to sensory afferent input have also been studied through stretch and cutaneous reflexes in populations with brain or spinal cord injury. Muscle spindle afferents can be activated via sudden involuntary changes in muscle length (i.e., stretch) to evoke reflexive muscle responses. This is commonly achieved through tendon taps/jerks or passive limb movements, which activate muscles spindles to send an excitatory volley to motoneurons (McNeil *et al.*, 2013). Unlike the highly synchronous afferent volley from the electrical stimulation of sensory fibres, which causes a brief EPSP in motoneurons, stretch reflexes like the tendon tap/jerk cause a more dispersed volley comprising multiple discharges and is longer in duration (Burke *et al.*, 1983). SSRIs increase the amplitude of stretch reflexes, and 5-HT$_2$ receptor antagonism decreases stretch reflexes in individuals with chronic spinal cord injury (Wei *et al.*, 2014). Moreover, trains of electrical stimuli can be applied to the skin at the foot to evoke cutaneous reflexes in lower-limb muscles, and are enhanced by single dose SSRI administration but reduced by 5-HT$_2$ antagonism and 5-HT$_{1B/1D}$ agonism in chronic spinal cord injured individuals (D'Amico *et al.*, 2013a; D'Amico *et al.*, 2013b; Murray *et al.*, 2010). It should be noted that even weak excitatory input can cause a long-lasting involuntary firing of motoneurons in chronic spinal cord injured subjects (D'Amico *et al.*, 2014). This motoneuronal hyperexcitability is partly due to plastic changes in monoaminergic receptors, such as the 5-HT$_2$ receptor, which becomes constitutively active, facilitates PICs and augments muscle spasticity (D'Amico *et al.*, 2013b; Murray *et al.*, 2010). Hence, individuals with chronic spinal cord injury are likely very sensitive to 5-HT drug interventions, and it is unlikely that a similar magnitude of 5-HT drug mediated changes in reflex amplitudes would be observed in intact human participants with a preserved descending control of motoneurons. In individuals with stroke, single dose SSRI administration increases paretic limb stretch reflexes (Gourab *et al.*, 2015), perhaps implicating 5-HT in the pathophysiology of post-stroke spasticity.

Vibration reflexes also cause the reflexive recruitment of motoneurons, whereby motoneurons can be activated repetitively via monosynaptic input from peripheral afferents when a muscle or tendon is subjected to sustained vibration (Monjo and Shemmell, 2020). Although assessed in resting muscle, and hence under conditions with weak serotonergic drive to the spinal cord (Jacobs *et al.*, 2002; Veasey *et al.*, 1995), SSRIs enhance the magnitude of vibratory reflexes in healthy subjects (Wei *et al.*, 2014), indicating a net facilitatory effect of 5-HT release on resting motoneurons. Vibration will provide a sustained ionotropic input to motoneurons, which allows the activation of slowly activating PICs, so 5-HT drug effects on muscle responses to vibration may approximate 5-HT effects on PICs for human motoneurons.

### 3. The noradrenaline system

*3.1 Noradrenaline increases the size of muscle responses to motor cortical stimulation*

Compared to the number of TMS experiments investigating 5-HT drugs, there were many more examining NA active drugs (Berger *et al.*, 2018; Boroojerdi *et al.*, 2001; Buchmann *et al.*, 2010; Buchmann *et al.*, 2007; Buchmann *et al.*, 2006; Butefisch *et al.*, 2002; Foster *et al.*, 2006; Gilbert *et al.*, 2006a; Gilbert *et al.*, 2006b; Gilbert *et al.*, 2007; Herwig *et al.*, 2002; Hoeppner *et al.*, 2008; Ilic *et al.*, 2003; Kesar *et al.*, 2017; Kirschner *et al.*, 2003; Klass *et al.*, 2016; Klass *et al.*, 2012; Klass *et al.*, 2018; Korchounov *et al.*, 2003; Korchounov and Ziemann, 2011; Kratz *et al.*, 2009; Kuo *et al.*, 2017a, b; Lange *et al.*, 2007; Meintzschel and Ziemann, 2006; Moll *et al.*, 2003; Moll *et al.*, 2000; Nitsche *et al.*, 2004; Ozdag *et al.*, 2010; Plewnia *et al.*, 2001; Plewnia *et al.*, 2004; Plewnia *et al.*, 2002; Sawaki *et al.*, 2002; Sawaki *et al.*, 2003; Schneider *et al.*, 2011; Sczesny-Kaiser *et al.*, 2014; Ziemann *et al.*, 2002; Zittel *et al.*, 2007). Hence, there has been considerable interest regarding noradrenergic transmission within motor pathways, and how NA drugs affect muscle responses to TMS of the motor cortex in humans.

Across the NA drug studies that we identified; it was consistently demonstrated that NA enhancing drugs increase the size of MEPs (Table 2). This indicates that NA has a net facilitatory effect on human motor cortical and/or spinal motoneuronal excitability. Importantly, in healthy participants, NA reuptake inhibitors (NRIs) that selectively increase the synaptic concentration of NA increase resting MEP amplitude (Kuo *et al.*, 2017b; Plewnia *et al.*, 2004; Plewnia *et al.*, 2002; Sczesny-Kaiser *et al.*, 2014), indicating that increases in motor system excitability can be caused by NA alone and not merely in combination with reuptake inhibition or facilitation of other neurotransmitter systems. For example, some studies indicate that noradrenaline dopamine reuptake inhibitors (NDRIs) and amphetamine also increase resting MEPs in healthy subjects (Boroojerdi *et al.*, 2001; Ilic *et al.*, 2003), i.e., drugs with strong dopaminergic effects in addition to noradrenergic. It also appears that both single dose and chronic administration of an NRI both increase resting MEP amplitude, and do not have an opposite direction of effect depending on the number of administrations (i.e., longer term use, as seen with SSRIs).

Antagonism of Gi/o coupled $\alpha_2$ receptors increase resting MEP amplitude (Plewnia *et al.*, 2001). NA producing neurons express $\alpha_2$ receptors, and these pre-synaptic $\alpha_2$ receptors reduce the release of NA through a negative feedback loop (Philipp *et al.*, 2002). Hence, it seems that $\alpha_2$ receptor antagonism impedes this negative feedback mechanism to promote the release of NA, which is why $\alpha_2$ antagonists and NRIs both increase MEP amplitude and activating $\alpha_2$ receptors with an agonist (which will reduce NA release) has the opposite effect and reduces resting MEP size (Korchounov *et al.*, 2003). Noradrenergic effects on the MEP are not limited to the $\alpha_2$ receptor, as antagonism of the Gq coupled $\alpha_1$ receptor has also been shown to increase the size of resting MEPs (Korchounov and Ziemann, 2011). Yet, MEPs were not affected by $\beta$ receptor antagonists in any study we found (Nitsche *et al.*, 2004; Sawaki *et al.*, 2003).

Overall, motor threshold was not affected by NA drugs (Boroojerdi *et al.*, 2001; Buchmann *et al.*, 2010; Buchmann *et al.*, 2007; Buchmann *et al.*, 2006; Butefisch *et al.*, 2002; Foster *et al.*, 2006; Gilbert *et al.*, 2006a; Gilbert *et al.*, 2007; Hoeppner *et al.*, 2008; Ilic *et al.*, 2003; Kesar *et al.*, 2017; Kirschner *et al.*, 2003; Korchounov *et al.*, 2003; Kratz *et al.*, 2009; Kuo *et al.*, 2017b; Lange *et al.*, 2007; Meintzschel and Ziemann, 2006; Moll *et al.*, 2003; Moll *et al.*, 2000; Ozdag *et al.*, 2010; Plewnia *et al.*, 2001; Plewnia *et al.*, 2004; Plewnia *et al.*, 2002; Sawaki *et al.*, 2002; Sawaki *et al.*, 2003; Schneider *et al.*, 2011; Sczesny-Kaiser *et al.*, 2014; Ziemann *et al.*, 2002), except for a single study showing a decrease in RMT with NRI administration (Herwig *et al.*, 2002). Thus, like the 5-HT system, this indicates that the NA system has minimal effects on (TMS-activated) lower threshold cortical circuits that project to motoneurons and/or the excitability of lower-threshold motoneurons.

*3.2 Noradrenaline increases intracortical facilitation but reduces inhibition*
Under conditions of pharmacologically heightened NA neurotransmission, the abovementioned increases in the size of MEPs could be the consequence of increased intracortical facilitation and/or a decreased intracortical inhibition. Several studies, as can be seen in Table 2, indicate that noradrenergic enhancing drugs simultaneously increase ICF and decrease SICI in healthy individuals (Boroojerdi *et al.*, 2001; Gilbert *et al.*, 2006a; Herwig *et al.*, 2002; Ilic *et al.*, 2003; Kirschner *et al.*, 2003; Kuo *et al.*, 2017b; Moll *et al.*, 2003; Plewnia *et al.*, 2004; Plewnia *et al.*, 2002; Sczesny-Kaiser *et al.*, 2014). Hence, noradrenergic neurotransmission appears to both augment intracortical facilitatory activity and suppress inhibition to increase motor cortical excitability.

With respect to ICF, a subthreshold facilitatory circuit may be augmented by NA. Specifically, enhanced noradrenergic transmission with drug administration perhaps increases the recruitment/output of excitatory interneurons and/or increases the responsiveness of corticospinal neurons to excitatory input. The $\alpha_2$ receptor can be linked to NA effects on ICF, whereby blocking the $\alpha_2$ receptor increases ICF (Plewnia *et al.*, 2001), and activation of the $\alpha_2$ receptor with an agonist reduces ICF (Korchounov *et al.*, 2003). Perhaps, as the pre-synaptic $\alpha_2$ receptor acts to inhibit NA release (Philipp *et al.*, 2002), drugs that target this receptor are simply controlling the availability of NA in the synapse and ICF effects are instead mediated by post-synaptic receptors. Yet, $\alpha_1$ and $\beta$ receptor antagonists do not affect ICF (Sawaki *et al.*, 2003). Regarding SICI, enhanced NA activity could reduce the output or efficacy of inhibitory cortico-cortical projections that act on corticospinal projections to motoneurons. An agonist for the $\alpha_2$ receptor increases SICI (Korchounov *et al.*, 2003), possibly by reducing NA release as explained above. Like $\alpha_1$ and $\beta$ antagonist effects on ICF, $\alpha_1$ and $\beta$ receptor antagonists do not affect SICI (Sawaki *et al.*, 2003).

The administration of NRIs and NDRIs have been shown to enhance SICI, and hence estimates of intracortical inhibition, in individuals with attention-deficit/hyperactivity disorder (ADHD), albeit paired pulse TMS results are not entirely consistent between studies involving this population (Buchmann *et al.*, 2007; Gilbert *et al.*, 2006b; Gilbert *et al.*, 2007; Moll *et al.*, 2000; Schneider *et al.*, 2011). LICI is reduced with chronic NDRI administration in healthy participants (Buchmann *et al.*, 2010), but LICI is increased with chronic NDRI administration in individuals with ADHD (Buchmann *et al.*, 2007). The TMS-induced silent period is not affected by any NA drug in healthy participants or in individuals with ADHD (Hoeppner *et al.*, 2008; Ilic *et al.*, 2003; Klass *et al.*, 2016; Klass *et al.*, 2012; Klass *et al.*, 2018; Korchounov *et al.*, 2003; Moll *et al.*, 2003; Moll *et al.*, 2000; Ozdag *et al.*, 2010; Sawaki *et al.*, 2003). Considering that NA drugs affect SICI and LICI but not the TMS-induced silent period, this could indicate that the inhibitory mechanisms or circuits that condition a MEP are likely distinct to inhibitory mechanisms causing the TMS-induced silent.

*3.3 Noradrenergic effects on spinal motoneurons*

We identified several studies examining NA drug effects on motoneuron excitability (Boroojerdi *et al.*, 2001; Brunia, 1972, 1979; De and Richens, 1974; Ilic *et al.*, 2003; Klass *et al.*, 2016; Klass *et al.*, 2018; Korchounov *et al.*, 2003; Mai, 1978; Mai and Pedersen, 1976; Phillips *et al.*, 1973; Plewnia *et al.*, 2001; Plewnia *et al.*, 2002; Udina *et al.*, 2010; White and Richens, 1974). Although we did not find one instance where cervicomedullary stimulation was used to assess human motoneuron excitability changes after the administration of NA drugs (Table 2), drugs that enhance NA activity (i.e., amphetamines, NRIs) increase the size of H-reflexes and stretch reflexes in healthy participants (Brunia, 1972; Klass *et al.*, 2018; Phillips *et al.*, 1973). This indicates that NA increases human motoneuron excitability to sensory afferent input.

The post-synaptic facilitatory effects of NA on human motoneuron excitability are probably mediated by the $\alpha_1$ receptor, as blocking this facilitatory receptor reduces stretch reflexes in healthy participants (De and Richens, 1974; Phillips *et al.*, 1973) and reduces both H- and stretch reflexes in individuals with brain and spinal injuries (Mai, 1978; White and Richens, 1974). Given that the enhanced availability of NA is non-selective to adrenergic receptors and likely also causes the augmented stimulation of the inhibitory $\alpha_2$ receptor located on sensory afferents, which reduces sensory afferent transmission to motoneurons (D'Amico *et* *al.*, 2014; Rank *et al.*, 2011), α$_1$ receptor activation appears to override these inhibitory influences and has a stronger influence on reflex amplitudes. Vibratory reflexes are also reduced by an α$_1$ receptor antagonist (Mai, 1978).

Activation of the α$_2$ receptor with an agonist reduces the size of stretch reflexes (De and Richens, 1974). There are two possibilities that could explain how α$_2$ receptor activation reduces the size of muscle responses to afferent input. As explained above, the reduction in stretch reflexes with α$_2$ receptor activation could be a consequence of reduced sensory transmission to motoneurons (D'Amico *et al.*, 2014; Rank *et al.*, 2011). On the other hand, activation of pre-synaptic inhibitory α$_2$ auto receptors on NA releasing neurons could reduce background release of NA into the synapse (Philipp *et al.*, 2002), meaning NA would be less able to activate post-synaptic α$_1$ receptors at motoneurons that appear to enhance motoneuron excitability.

Interestingly, an isolated study indicates that β receptor antagonism may increase the size of stretch reflexes in healthy participants (Brunia, 1979), suggesting that this receptor augments the excitability of a component/s of the spinal sensorimotor loop. Yet, NA drugs do not affect F-waves (Boroojerdi *et al.*, 2001; Ilic *et al.*, 2003; Korchounov *et al.*, 2003; Plewnia *et al.*, 2001; Plewnia *et al.*, 2002), indicating that the recurrent discharge of human motoneurons to antidromic activation is not affected by this neuromodulator. It is important to note, however, that F-waves may not always be a sensitive test to identify changes in motoneuron excitability (Balbi, 2016; Lin and Floeter, 2004), so it is not unusual for F-waves to remain unaffected by interventions that affect other measures of motoneuron excitability (e.g., H-reflexes or CMEPs).

**4. Drugs with combined serotonin and noradrenaline effects support a role for monoamines in the excitability of motor pathways**

Several studies that used a drug intervention targeting both NA and 5-HT reuptake or used antagonists with a strong affinity for both 5-HT and NA receptors were identified (Basmajian and Szatmari, 1955; Brunia, 1973; D'Amico *et al.*, 2013b; De and Richens, 1974; Foster *et al.*, 2006; Li *et al.*, 2018; Li and Morton, 2020; Lissemore *et al.*, 2021; Metz *et al.*, 1982; Munchau *et al.*, 2005; Stern *et al.*, 1968). The results of these studies are summarised in Table 3. A notable finding was that combined α$_2$ and 5-HT$_2$ receptor antagonism lengthens the TMS-induced silent period in healthy participants without affecting motor threshold, unconditioned and conditioned MEPs (Munchau *et al.*, 2005). Considering that a more selective 5-HT$_2$ antagonist also lengthens the TMS-induced silent period (Thorstensen *et al.*, 2021), but a drug selectively targeting the α$_2$ receptor has no effect on the same outcome (Korchounov *et al.*, 2003), the 5-HT$_2$ receptor effects of the drug could explain this result, which provides further support for a role of 5-HT in the enhancement of intracortical inhibition. Nevertheless, combined α$_2$ and 5-HT$_2$ receptor antagonism lowers AMT in individuals with co-occurring depression and epilepsy but has no effect on single and paired pulse MEPs or the TMS-induced silent period in this population (Munchau *et al.*, 2005). At the level of the spinal cord, combined α$_1$ and 5-HT$_2$ antagonism tends to reduce the size of H- and stretch reflexes (Basmajian and Szatmari, 1955; Brunia, 1973; De and Richens, 1974; Stern *et al.*, 1968), providing further support that 5-HT and NA acts via α$_1$ and 5-HT$_2$ receptors to increase motoneuron excitability in humans.

**5. Behavioural inferences**

*5.1 What is the neurobiological relevance of neuromodulation within motor pathways?*

We were interested in collating studies examining the effects of monoaminergic drugs on the excitability of the human motor cortex and motoneurons, as these studies best indicate whether a particular monoamine system or receptor can be implicated in motor cortex and/or motoneuron function. Indeed, human drug studies tell us that both 5-HT and NA neurotransmission greatly affect the excitability of the motor system at both cortical and spinal levels. Yet, considering that most of the studies included in this review involved resting measurements of motor cortical and/or motoneuronal excitability, i.e., when participants are sitting in a controlled lab environment in a relaxed state, these studies do not provide much insight regarding the behaviours that cause 5-HT and NA release in the human brain and spinal cord. Therapeutic doses of monoaminergic drugs simply mimic or 'knockout' the normal physiological effects exhibited by 5-HT and NA to enhance or reduce the effects of a particular transmitter system (compared to a condition without a drug).

Animal studies can help elucidate the relevance of monoaminergic neurotransmission and the behaviours that cause monoamine release to the motor system. Extracellular recordings from the brainstem show that the firing rates of descending raphe 5-HT producing neurons correlate with the speed of locomotion in cats (Jacobs *et al.*, 2002; Veasey *et al.*, 1995), whereby faster locomotion results in faster firing rates and 5-HT release to the spinal cord and motoneurons. Considering this, and the results of our review indicating a facilitatory effect of 5-HT on human motoneuron excitability (under conditions without muscle fatigue, Table 1), it seems likely that a primary function of the descending 5-HT system is to increase motoneuron excitability during strong motor behaviours requiring significant output from the corticospinal pathway and motoneurons (e.g., during movements requiring strong motor outflow to muscle). Under these conditions, it is entirely likely that the purpose of large 5-HT release to the cord is to increase motoneuron excitability and make it easier for descending ionotropic pathways to recruit motor units and maintain their discharge with minimal excitatory drive.

The most prominent collection of NA cells forms the pontine locus coeruleus (Berridge and Waterhouse, 2003; Poe *et al.*, 2020). The tonic discharge activity of locus coeruleus noradrenergic neurons is scaled to the level of arousal, whereby there are low to moderate firing rates during quiet waking but fast firing rates upon conditions that require high arousal (Berridge *et al.*, 2012; Berridge and Waterhouse, 2003). This indicates that NA release to the brain and spinal cord is augmented during states of high alertness, stress, or anxiety, i.e., during behaviours that are associated with a heightened activation of the sympathetic nervous system. Perhaps, as NA enhancing drugs increase the excitability of the human motor cortex and motoneurons (Table 2), and NA release is large with heightened sympathetic activity, we suggest that NA circuits are needed to strengthen the output of the motor cortex and motoneurons under conditions of hyperarousal (colloquially labelled 'fight or flight' conditions). From an evolutionary standpoint, the gain of the motor system may be augmented to quickly activate muscle and escape a predator when we want to 'flight', and more powerfully activate muscle when we need to 'fight'.

Notably, both 5-HT and NA producing cells are quiescent during rapid eye movement (REM) sleep (Berridge and Waterhouse, 2003; Jacobs *et al.*, 2002), a stage of sleep characterised by reductions in muscle tone and a complete suppression of motor output. Consistent with our above suggestions, this makes sense. High levels of neuromodulation of the motor system and the enhanced excitability of motoneurons are not needed at complete rest when muscles do not need to be engaged. Alternatively, in a waking state, 5-HT and NA projections could serve the combined function of a tonic gain regulator for the motor system to maintain background muscle tone, and 5-HT (for strong motor outflow) and NA (with high stress situations) release upregulated when required.

*5.2 Future directions*

There needs to be a better effort in future human drug studies to explore the behaviours that cause 5-HT and NA release in the brain and spinal cord, and the resultant functional effect/s that monoamine release has on the output of motor pathways. For example, considering the motor activity-dependent nature for 5-HT release, a 5-HT$_2$ receptor antagonist might be employed to block the facilitatory effects of 5-HT across different intensities of isometric muscle contractions or locomotor tasks that manipulate the magnitude of 5-HT release. If 5-HT release is also coupled to motor activity in humans as it appears to be in animals, it might be hypothesised that stronger motor activities would make the effects of the drug on muscle responses to neurostimulation more noticeable. Similarly for the NA system, an $\alpha_1$ receptor antagonist might be combined with virtual reality interventions that acutely heighten stress or arousal to investigate the effects of NA release.

## 6. Conclusions

The key take home message from this review is that the monoaminergic neuromodulators 5-HT and NA have strong effects on the excitability of the human motor cortex and motoneurons. When TMS is used to test the excitability of motor cortical circuits, drugs that enhance 5-HT activity tend to reduce measures of facilitation and augment measures of inhibition, indicating that 5-HT acts to reduce intracortical excitability. Conversely, NA promoting drugs have the opposite effects, markedly increasing motor cortical excitability across many studies and populations, whereby the pharmacological enhancement of NA activity increases paired-pulse TMS measures of intracortical facilitation and reduces measures of inhibition. Considering that TMS of the motor cortex assesses the excitability of the entire corticospinal-motoneuronal system from the cortex to muscle, 5-HT and NA drugs could have acted at motoneurons to cause these excitability changes. Indeed, drug studies indicate that 5-HT has powerful effects on human motoneurons, whereby 5-HT enhancing drugs increase motoneuron excitability through 5-HT$_2$ receptor activation, but activation of 5-HT$_1$ receptors may also reduce human motoneuron excitability under specific conditions (i.e., with fatiguing muscle contractions) or reduce sensory transmission to motoneurons. Likewise, NA enhancing drugs increase the excitability of human motoneurons to synaptic input, and it is likely that the facilitatory effects of NA on human motoneuron excitability are mediated by $\alpha_1$ receptors.

Hopefully this review brings to attention a role for endogenously released monoamine neuromodulators in the function of motor circuits in the human brain and spinal cord. More work needs to be done to understand how these monoamines control the excitability of the human motor cortex and motoneurons in health and disease.

**Acknowledgements**

There are no acknowledgements to declare.

**Funding**

This research did not receive any specific grant from funding agencies in the public, commercial, or not-for-profit sectors.

**Competing interests**

None to declare.

**Author contributions**

Jacob Thorstensen: Conceptualisation, literature review and creation of tables, writing, editing. Tyler Henderson: Conceptualisation, editing and feedback. Justin Kavanagh: Conceptualisation, editing and feedback**.**

**Table 1.** Results of studies assessing effects of serotonin drugs on motor cortical and motoneuronal excitability.

| Corticospinal excitability with TMS | | RMT | AMT | Resting MEP | Active MEP | ICF | SICI | LICI | Silent period |
|---|---|---|---|---|---|---|---|---|---|
| Population | Drug class | | | | | | | | |
| Healthy | SSRI | ○ ○ ○ ○ ○ ▲ | ○ ○ ○ | ○ ○ ○ ○ ○ ▲ ▲ | ○ ○ | ▼ ▼ ○ ○ | ○ ○ ▲ ▲ | | ○ ▲ ▲ ▲ |
| | SSRI (chronic) | ○ ○ | ○ | ▼ ○ | | | ▲ | ○ | ○ |
| | 5-HT2 antagonist | | ○ | | ▼ | | | | ▲ |
| | 5-HT1B/1D agonist | ○ | | ○ | ○ | ○ | ▼ | | ○ |
| Depression | SSRI | ▲ | | ○ ▲ | ▲ | ▼ | ▲ | | ○ |
| | SSRI (chronic) | | | | | ○ | ○ | ▼ | |
| Stroke | SSRI (chronic) | Affected ○ Unaffected ▲ | | Affected ○ Unaffected ○ ○ | | Unaffected ▲ | Affected ○ Unaffected ▲ ▼ | | |
| Stuttering | SSRI (chronic) | ○ | ○ | | | | | | ▼ |

| Motoneuron excitability | | CMEP | H-reflex | Stretch reflex | Vibratory reflex | Cutaneous reflex | F-waves | Estimates of PICs | Silent period |
|---|---|---|---|---|---|---|---|---|---|
| Population | Drug class | | | | | | | | |
| Healthy | SSRI | | | | ▲ | | ○ ○ ○ ▼* | ▲ | ▲ |
| | SSRI (chronic) | | | | | | ○ | | |
| | 5-HT2 antagonist | ○ | | | | | ▼ | ▼ | |
| | 5-HT1B/1D agonist | | ▼ | | | | | | |
| | 5-HT1A agonist | ▼ | | | | | ▼ | | |
| Depression | SSRI | | | | | | ○ | | |
| | SSRI (chronic) | | | | | | ○ | | |
| Stroke | SSRI | | | Affected ▲ | | | | | |
| | 5-HT2 antagonist (chronic) | | | Affected ○ | | | | | |
| SCI | SSRI | | | ▲ | | ▲ | | | |
| | 5-HT2 antagonist | | | ▼ | | ▼ ▼ | | ▼ | |
| | 5-HT1B/1D agonist | | | ▼ | | ▼ | | | |

All studies involved single dose administration of a drug unless indicated as chronic (administration over weeks or months). ○ = no change in outcome, ▼ = decrease, ▲ = increase. For stroke studies, effects have been dichotomised for the affected and unaffected hemispheres (TMS studies), or affected and unaffected limbs (motoneuron excitability studies). 5-HT = 5-hydroxytryptamine/serotonin, AMT = active motor threshold, CMEP = cervicomedullary motor evoked potential, ICF = intracortical facilitation, LICI = long-interval intracortical inhibition, MEP = motor evoked potential, PICs = persistent inward currents, RMT = resting motor threshold, SCI = spinal cord injury, SICI = short-interval intracortical inhibition, SSRI = selective serotonin reuptake inhibitor, TMS = transcranial magnetic stimulation. For references, see main text.
*For F-waves, one study found a SSRI-mediated decrease only for fatigued muscle, but all other studies report drug effects in an unfatigued state.

**Table 2.** Results of studies assessing effects of noradrenaline drugs on motor cortical and motoneuronal excitability.

| Corticospinal excitability with TMS | | RMT | AMT | Resting MEP | Active MEP | ICF | SICI | LICI | Silent period |
|---|---|---|---|---|---|---|---|---|---|
| Population | Drug class | | | | | | | | |
| Healthy | Amphetamine/Amphetamine prodrug | ○ ○ ○ ○ ○ | | ▼ ○ ○ ○ ○ ▲ | | ○ ▲ | ○ | | |
| | NRI | ▼ ○ ○ ○ ○ ○ ○ | ○ ○ | ○ ○ ▲ ▲ ▲ ▲ | ○ ○ ▲ | ▲ ▲ ▲ ▲ ▲ ▲ | ▼ ▼ ▼ ○ ○ | | ○ ○ ○ |
| | NRI (chronic) | ○ ○ | ○ | ▲ | | ○ ▲ | ○ ▼ | | |
| | NDRI | ○ ○ ○ ○ ○ ○ ○ | ○ ○ ○ ○ | ○ ○ ○ ○ ▲ | ○ | ○ ▲ ▲ ▲ | ▼ ▼ ○ ○ ▲ | | ○ ○ ○ |
| | NDRI (chronic) | ○ | | ▲ | | ○ ▲ | ○ ▼ | ▼ ○ | |
| | α2 antagonist | ○ | | ▲ | | ▲ | ○ | | |
| | α2 agonist | ○ | ○ | ▼ | | ▼ | ▲ | | ○ |
| | α1 antagonist | ○ ○ | ○ | ○ ○ ▲ | | ○ | ○ | | ○ |
| | β antagonist | ○ | ○ | ○ ○ | | ○ | ○ | | ○ |
| ADHD with Tourette Syndrome | NRI | ○ | ○ | ○ | | ▲ | ○ ▲ | | |
| | NRI (chronic) | ○ | ○ | ○ | | ○ | ○ | | |
| | NDRI | | | | | | ▲ | | |
| ADHD | NRI | | | | | | ▼ | | |
| | NDRI | ○ ○ | ○ | | | ○ | ▼ ▲ | | ○ ○ |
| | NDRI (chronic) | ○ ○ ○ ○ | ○ | ○ ○ ○ | | ○ ▲ | ▲ ▲ | ▲ | ○ |
| Stroke | NRI | | | Affected ○<br>Unaffected ○ | | | | | |

| Motoneuron excitability | | CMEP | H-reflex | Stretch reflex | Vibratory reflex | Cutaneous reflex | F-waves | Estimates of PICs | Silent period |
|---|---|---|---|---|---|---|---|---|---|
| Population | Drug class | | | | | | | | |
| Healthy | Amphetamine/Methylamphetamine | | ○ ▲ | ▲ ▲ | | | ○ | ▲ | |
| | NRI | | ○ ○ ▲ | | | | ○ | | |
| | NDRI | | | | | | ○ | | ○ |
| | α2 antagonist | | | | | | ○ | | |
| | α2 agonist | | | | ▼ | | ○ | | ○ |

| | | | | | |
|---|---|---|---|---|---|
| | α1 antagonist | | ○ | ▼▼ | |
| | β antagonist | | ○ | ○ ▲ | |
| Brain/spinal cord lesions (otherwise undefined) | α1 antagonist | ▼ | ▼▼ | ▼ | ○ |
| | β antagonist | | ○ | ○ | |

All studies involved single dose administration of a drug unless indicated as chronic (administration over weeks or months). ○ = no change in outcome, ▼ = decrease, ▲ = increase. For the single stroke study, effects have been dichotomised for the affected and unaffected hemispheres. ADHD = attention-deficit/hyperactivity disorder, AMT = active motor threshold, CMEP = cervicomedullary motor evoked potential, ICF = intracortical facilitation, LICI = long-interval intracortical inhibition, MEP = motor evoked potential, NDRI = noradrenaline dopamine reuptake inhibitor, NRI = noradrenaline reuptake inhibitor, PICs = persistent inward currents, RMT = resting motor threshold, SICI = short-interval intracortical inhibition, TMS = transcranial magnetic stimulation. For references, see main text.

**Table 3.** Results of studies assessing effects of drugs with combined serotonin and noradrenaline properties on motor cortical and motoneuronal excitability.

| Corticospinal excitability with TMS | | RMT | AMT | Resting MEP | Active MEP | ICF | SICI | LICI | Silent period |
|---|---|---|---|---|---|---|---|---|---|
| Population | Drug class | | | | | | | | |
| Healthy | SNRI | ○ | | ○ | | | | | |
| | SNRI (chronic) | ○ | | ○ | | | | | |
| | α2/5-HT2 antagonist | ○ | ○ | ○ | ○ | ○ | ○ | | ▲ |
| Depression | SNRI (chronic) | | | | | ○ | ○ | | ○ |
| Depression with epilepsy | α2/5-HT2 antagonist | ○ | ▼ | ○ | ○ | ○ | ○ | | ○ |
| | α2/5-HT2 antagonist (chronic) | ○ | ○ | ○ | ○ | ○ | ○ | | ○ |
| Stroke | SSRI or SNRI (chronic) | Affected ○ Unaffected ○ | | Affected ○ Unaffected ○ | | | | | |

| Motoneuron excitability | | CMEP | H-reflex | Stretch reflex | Vibratory reflex | Cutaneous reflex | F-waves | Estimates of PICs | Silent period |
|---|---|---|---|---|---|---|---|---|---|
| Population | Drug class | | | | | | | | |
| Healthy | α1/5-HT2 antagonist | | ▼ ○ ○ | ▼ ▼ | | | | ▼ | |
| SCI | α1/5-HT2 antagonist | | | | | ○ | | | |
| Brain injury/SCI | α1/5-HT2 antagonist | | | ▼ | | | | | |
| Schizophrenia | α1/5-HT2 antagonist | | ○ | | | | | | |

All studies involved single dose administration of a drug unless indicated as chronic (administration over weeks or months). ○ = no change in outcome, ▼ = decrease, ▲ = increase. For the single stroke study, effects have been dichotomised for the affected and unaffected hemispheres. AMT = active motor threshold, CMEP = cervicomedullary motor evoked potential, ICF = intracortical facilitation, LICI = long-interval intracortical inhibition, MEP = motor evoked potential, PICs = persistent inward currents, RMT = resting motor threshold, SCI = spinal cord injury, SICI = short-interval intracortical inhibition, SNRI = serotonin noradrenaline reuptake inhibitor, SSRI = selective serotonin reuptake inhibitor, TMS = transcranial magnetic stimulation. For references, see main text.